\documentclass[12pt]{iopart}
\usepackage{amssymb}
\usepackage{epsfig} 
\usepackage{graphicx} 
\usepackage{epstopdf}
\usepackage{float}

\def \sq{\tilde q}
\def \gl{\tilde g}
\def\t1{\tilde t_1}
\def \CH{{\tilde\chi}^{\pm}}
\def \N0{\tilde\chi^0}
\def \PMET{p{\!\!\!/}_T}

\def \lumi{{\rm fb}^{-1}}
\def\bul{\bullet}
\begin{document}

\title{Higgs signal in Chargino-Neutralino production at the LHC}

\author{Diptimoy Ghosh$^a$,Monoranjan Guchait$^b$,Dipan Sengupta$^b$}
\address{$^a$ Department of Theoretical Physics, Tata Institute of 
Fundamental Research, Homi Bhabha Road, Mumbai, India}
\ead{diptimoyghosh@theory.tifr.res.in}

\address{$^b$ Department of High Energy Physics, Tata Institute of Fundamental 
Research,Homi Bhabha Road, Mumbai, India}
\ead{guchait@tifr.res.in,  dipan@tifr.res.in}


\begin{abstract}
We have analyzed the prospect of detecting a Higgs signal in 
mSUGRA/cMSSM based Supersymmetric (SUSY) model
via chargino-neutralino($\CH_1\N0_2$) production at 8 TeV
and 14 TeV LHC energies. 
The signal is studied in the $\ell + b \bar b + \PMET$ channel 
following the decays, $\CH_1 \to W^\pm \N0_1$, $\N0_2 \to \N0_1 h$ and 
$h \to b \bar b$. In this analysis reconstruction of the Higgs mass out of
two b-jets plays a very crucial role in determining the signal to 
background ratio. We follow two techniques to reconstruct the 
Higgs mass: (A) adding momenta of two identified b-jets, (B) jet 
substructure technique. In addition, imposing a certain set of 
selection cuts we observe that 
the significance is better for the method (B). We find that a  signal
can be observed for the Higgs mass  $\sim$ 125 GeV 
with an integrated luminosity 100~fb$^{-1}$ for both 
8~TeV and 14~TeV LHC energies.
\end{abstract}


\section{Introduction}
The quest for the  Higgs boson is one 
of the high priority programme of the LHC 
experiment.  Recently, both  ATLAS and CMS collaborations have published 
 preliminary
results on Higgs searches accumulating data of integrated luminosity about 5$\lumi$ each
at 8~TeV LHC energy.
They have constrained the light Higgs mass 
within the range of 122 - 131 GeV at 95\% C.L. which is
 \cite{lhc} consistent
with the prediction based on electro-weak precision 
measurements \cite{pdg}. 
However, interestingly both the groups have also reported the discovery of a 
 standard model(SM) like
Higgs boson with mass $\sim 125 $ GeV~\cite{lhc}. Further investigations are
required to confirm that it is indeed the Higgs boson predicted by
the SM.

In this paper we investigate the implications of Higgs searches
mentioned above assuming that the observed boson is indeed the 
Higgs boson. As we know, many models beyond the SM also predict 
the existence of a Higgs particle. For instance, the Minimal 
Supersymmetric Standard Model(MSSM) contains five Higgs bosons,
 two CP even 
Higgs $h$ and $H$, one CP odd Higgs $A$ and two charged Higgs 
$H^\pm$. At the tree level, the masses of all the Higgs particles 
in the MSSM can be predicted in terms of the two parameters,
 the CP-odd 
Higgs mass $m_A$ and the ratio $\tan\beta$ of the vacuum expectation 
values of the two Higgs doublets. The mass of the lightest Higgs($m_h$) 
is bounded by $m_h \leq m_Z$~\cite{abdel2} at the tree level, but loop
corrections enhance this limit to $m_h \lesssim 140$~GeV\cite{abdel2}.
Notice that this theoretical upper limit is consistent with the present 
limits set on the Higgs mass by LHC experiments. Note that, in the 
decoupling regime  $m_A >> m_Z$, the lightest Higgs becomes SM like. 
Evidently, there is a correlation between $m_{h}$ and other sparticle 
masses(and hence other model parameters) 
because of loop effects. 
The dependence of $m_h$ on model parameters including recent Higgs mass
constraints 
are discussed
by a number of authors in the framework of constrained MSSM(cMSSM) or
the minimal supergravity(mSUGRA)~\cite{vbarger,abdel,pnath} model 
and also other variations of SUSY models~\cite{abdel}.

The mSUGRA model is described by four parameters, 
$m_0, m_{1/2}, A_0$(defined at the GUT scale), 
$\tan\beta$(defined at the Electroweak scale) 
and a sign, the sign of $\mu$. 
Here $m_0$ is the universal soft mass of the scalars, $m_{1/2}$ is the 
unified gaugino mass, 
$A_0$ is the universal trilinear coupling and $\mu$ is the supersymmetric 
Higgs(ino) mass 
parameter. The lightest Higgs mass is highly sensitive to $m_0, A_0$ 
and $\tan\beta$, 
as the square of the third generation 
squark mass matrix which contributes dominantly to the loop 
correction is controlled by these parameters. 
A detailed scan of parameter 
space shows that the current constraints on the Higgs mass from LHC experiments
is compatible with certain regions of parameter space in mSUGRA/cMSSM. 
For example, for low $m_0$($\le$4 TeV) case, to achieve 
$m_h \sim 125$~GeV, a high value of  
$A_0$ is required, whereas for high $m_0 \sim$ 4 TeV, 
one needs a moderate
value of $A_0$\cite{abdel,pnath}. 
As a consequence,  the parameter space in mSUGRA 
allowed by Higgs mass
constraints predict the masses of sfermions(squarks and sleptons)
to be of multi-TeV range. 
However, the mass of top squark($\tilde{t_{1}}$)
remains comparatively lighter because of 
mixing effects and is likely to be 
accessible within the LHC energy range along with 
gauginos(charginos and neutralinos) 
and gluinos~\cite{pnath}. It is worth mentioning here 
that from the negative results in direct searches at LHC, 
both ATLAS and CMS collaborations have excluded a region in the $m_0 - m_{1/2}$ plane 
imposing a limit, $m_{\gl} \gtrsim$ 1.2 TeV for $m_{\sq} \sim m_{\gl}$ case, 
and $m_{\gl} \gtrsim$ 800 GeV for $m_{\sq} >> m_{\gl}$ scenario\cite{lhcsusy}.

In this paper, we explore the detectability of Higgs signal 
in SUSY cascade decay chain which may enable us to confirm the existence 
of a SUSY Higgs. With this motivation we investigate the Higgs signal in 
chargino($\CH_1$) and second lightest neutralino($\N0_2$) pair production 
following the dominant decays, $\CH_1 \to \N0_1 W^{\pm}$ and 
$\N0_2 \to \N0_1 h$. Higgs signal in SUSY cascade decays has been studied 
previously in detail by the authors of Ref.\cite{asesh}.
It is well known that in hadron colliders, strongly interacting colored 
sparticles, $\gl$ and $\sq$ are produced copiously. 
The current exclusions by the LHC experiments from SUSY and Higgs 
searches in mSUGRA favor high $\tilde{q}$ and $\tilde{g}$ 
masses($m_{\gl}, m_{\sq}\sim$1 TeV ). For these ranges of $\gl$ and 
$\sq$ masses, the $\gl$ pair production is expected to dominate over the 
$\sq$ production. Eventually, the Higgs boson may arise in $\gl$ cascade 
decay chains involving heavy flavors, i.e., 
$\gl \to t b \CH_1, t \bar t \N0_1,t \bar t \N0_2$ and $\N0_2 \to \N0_1 h$.
We checked that the probability of finding Higgs events via $\gl$ pair 
production and its subsequent cascade decays is $\sim$ 1-3\%. Moreover, 
with the increase of $\gl$ and $\sq$ masses, strong production cross 
sections drop significantly ($\sim$ few fb) and electro-weak gaugino 
pair production takes over. In view of this fact, we consider $\CH_1\N0_2$ 
pair production to study the Higgs signal instead of the $\gl$ pair 
production. The detection of Higgs signal in this channel has not been 
studied before for 8 TeV LHC energy. 
It is to be noted that the $\CH_1\N0_2$ production is regarded to 
be a promising SUSY discovery channel through the clean trilepton signal. 
Recently, this channel has also received a lot of attention to probe
SUSY signal~\cite{mrena,tata} at the LHC due to the higher limits 
on $\gl$ and $\sq$ masses~\cite{lhcsusy}. 
Similar analysis has also been performed for LHC  
in \cite{barger} for 14 TeV energy. 
The Higgs production via $\t1$ production 
and its subsequent decays has also been discussed in \cite{heine}. 

In mSUGRA, at the GUT scale masses of all the gauginos are given 
by $m_{1/2}$ and at the electro-weak scale 
they are related  as $M_2 \simeq M_3/3 \sim m_{\gl}/3$ and $M_1 \simeq M_2/2$
because of renormalization group evolution(RGE). Here $M_1$, $M_2$ and $M_3$ 
are the U(1), SU(2) and SU(3) gaugino mass parameters respectively. 
%
%
To get a reasonable branching ratio for the decay $\N0_2 \to \N0_1 h$ 
we select parameter space where $|\mu|$, the Higgsino mass parameter 
is very large leading to $\CH_1$, $\N0_2$ and $\N0_1$ states gaugino 
dominated. Therefore, for very high values of $|\mu|$ 
(i.e., $|\mu| >> M_2, M_1$), $m_{\CH_1}, m_{\N0_2} \sim M_2 \sim M_{\gl}/3$.
Hence, in view of the current limit on $\gl$ mass, 
$m_{\CH_1}$ and $m_{\N0_2}$ are expected to be  around 
250~GeV or more. 
For the sake of presentation  of our results 
we select three
benchmark points(P1-P3) corresponding to progressively higher values of 
gaugino masses which are presented in Table ~\ref{table:table1}.
\begin{table}[t!]
\begin{center}
\begin{tabular}{l l l l l l l l l l }
\hline
& $m_{1/2}$ & $\mu$&  $m_h$ & $m_{\gl}$ & $m_{\sq}$ & $m_{\t1}$ & $m_{\N0_1}$
& $m_{\N0_2}$ & $m_{\CH_1}$ \\
\hline
P1 & 300 & 1541 & 122.4 & 865  & 3000 & 1305 & 133 & 265 & 265  \\
P2 & 380 & 1660 & 122.8 & 1046 & 3060 & 1335 & 168 & 332 & 332  \\
P3 & 450 & 1653 & 123.2 & 1200 & 3096 & 1370 & 198 & 390 & 390 \\
\hline
\end{tabular} 
\caption{ \label{table:table1} Masses of some of the sparticles for three benchmark points.
In all the cases $m_0=3000$, $\tan\beta$=30 and $A_0$=-4500. All mass  
units are in GeV.}
\end{center}
\end{table} 
For this region of parameter space, $\N0_2 \to \N0_1 h$ branching 
ratio(BR) is more than 
$ 80\%$ and the $h\to b \bar{b}$ BR is about 70\%. 
\\
In the next sections we discuss our simulation strategy for 
signal and backgrounds and then 
present our results. Finally we summarize our study in the last section.

\smallskip

\section{Signal and Background}
We investigate the Higgs signal in SUSY cascade decay,
$pp \to \CH_1 \N0_2 \to (\N0_1 W^{\pm})( \N0_1 h)$, leading to a final state
with a hard lepton($e,\mu$) from W decay and two b-jets from Higgs decay
and a large $\PMET$ due to the presence of $\N0_1$ and $\nu$, but without any 
additional jet. 
The identical final state may also come from ${\rm t \bar t}$, 
${\rm Wb\bar b}$, ${\rm Zb\bar b}$, $WZ$, $Wh$, $Zh$, $tb$, $tbW$ 
processes. Recall that
the $\CH_1 \N0_2$ pair production cross sections(C.S.) for our
considered parameter space are about 30-175 fb (LO) for 8 TeV
in contrast to background cross sections which vary from few picobarn(pb) to 
more than 100 pb. Thus a huge suppression
of background events is required to achieve 
a reasonable sensitivity, which is a 
challenging task. 
The added advantage is that the invariant mass constructed out of 
two b jets is expected to show a peak at the Higgs mass,
which can be exploited to identify the signal region.
Therefore, a good reconstruction of Higgs mass out of two b jets is one
of the crucial issue to be studied  in this analysis. 
In this paper we report about the 
simulation of signal and 
backgrounds adopting two methods for Higgs reconstruction. In the 
first method(Method A) we identify two b-jets out of 
all jets in the events and obtain the Higgs mass by  
calculating their invariant mass. 
In the second method(Method B) reconstruction of Higgs mass 
is performed by using the jet substructures which will be discussed later.
In this paper we present our results for both cases, method A and B.

In our simulation, events are generated using {\tt PYTHIA}\cite{pythia} 
for the signal and $t \bar t$, $WZ$, $Wh$, $Zh$ backgrounds whereas 
{\tt ALPGEN}\cite{alp} interfaced with {\tt PYTHIA} 
has been used for the generation of $tb$, $tbW$, $Wb\bar{b}$ and 
$Zb\bar{b}$ backgrounds.
We adopt MLM matching~\cite{mlm} to avoid double
counting while performing parton showering
after matrix element calculations in ALPGEN.
 We use {\tt FastJet} 
for jet reconstruction 
using built-in anti-$k_T$ algorithm 
with $\Delta R$=0.5 \cite{salam} in method A,
 whereas Cambridge-Aachen \cite{CA}
algorithm is used for method B. We use CTEQ6L parton distribution
function while calculating cross sections \cite{cteq}.
{\tt SuSpect} interfaced with {\tt SUSYHIT} is used to calculate
SUSY mass spectrum and corresponding branching ratios \cite{suspect}.

\begin{figure}[t]
\centering
\begin{tabular}{ll}
\includegraphics[height=2.80in,width=3.0in]{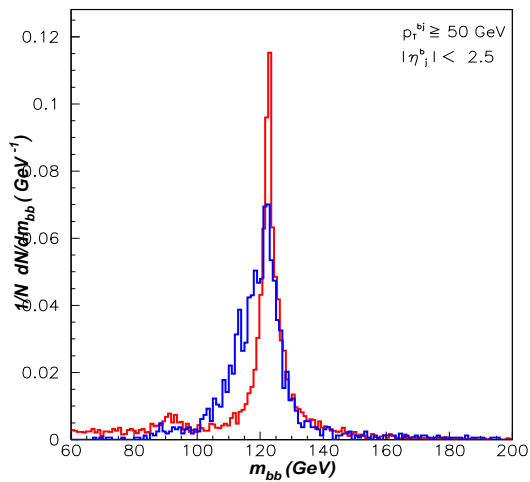} 
\end{tabular} 
\caption{\label{fig:fig1} The reconstructed Higgs mass for method A (blue) 
and method B (red) for $\rm \sqrt{s}=8~TeV$. The y-axis is normalized to unity.}
\end{figure}

We observe that use of Higgs mass reconstruction alone is not 
enough to eliminate backgrounds substantially. A certain set of selection 
cuts described below are necessary 
to reject backgrounds. 
\\
$\bullet$ Lepton : 
Leptons (e and $\mu$) are selected 
with $p_{T}^{\ell}\ge 20$~GeV and $|\eta|\le 2.5$. Isolation of leptons
are ensured by estimating the total transverse energy $p_T^{AC}\lesssim$20\% of
$p_T^{\ell}$, 
where $p_T^{AC}$ is the scalar sum of transverse energies 
of jets close to leptons satisfying
$\Delta R(\ell,j)\le$0.2. We veto events if there exists a second
lepton with a loose 
criteria of $p_{T}^{\ell}\ge 10$ GeV,
primarily to suppress top background.
\\
$\bullet$ Jets: Jets are selected using {\tt{FastJet}}~\cite{salam}
with a $p_T \ge$50~GeV and $|\eta|\le$3($|\eta|<$2.5 for method B). 
\\
$\bullet$ b-Jets: b like jets are identified by  performing a 
matching of jets  with b quarks assuming a matching cone
$\Delta R(b,j)=$0.5. 
In addition, we require that the matched b  jet transverse  momentum
 should have 
at least 80\% of the b quark transverse momentum. 
A proper method of b-tagging using displaced vertex
is beyond the scope of this analysis. Finally, we multiply by a  b-tagging
efficiency($\epsilon_b$) of 70\%~\cite{btag} for each b-tagging 
i.e $\epsilon_b^2$=0.5
for two b-tagged jets while estimating total event rates.
\\
$\bullet$ $p{\!\!\!/}_T$:
Missing transverse momentum is calculated out of of all visible 
stable particles. The $\PMET$ in the signal arises due to 
the massive $\N0_1$ and  $\nu$ as well, 
whereas in background events, it solely arises from $\nu$ in W decay.
Nevertheless, the hardness of $\PMET$ in signal is not significantly different 
than the large ${ t \bar t}$ background making it very difficult to 
distinguish the signal from the background.  
\\
$\bullet$ ${R_{T}^{b\bar{b}} }:$ We define a very robust variable
which is extremely efficient in eliminating backgrounds by huge 
fraction as discussed in a previous analysis~\cite{dipan}. It is 
defined as $ R_{T}^{b \bar b} = \frac{ p_T^{b_1} + p_T^{b_2}}{H_T}$, 
where the numerator is the scalar sum of $p_T$ of the two b-jets and 
$H_{T}$ is the scalar sum of $p_{T}$ of all jets passing our pre-selection 
criteria. We define this variable keeping in mind that in the signal process 
no hard jets are expected except two b-jets from Higgs decay. 
Of course, few jets may arise from initial and final state radiations, 
but the number of such jets with $p_T\ge$ 50 GeV is expected to be low. 
Hence $ R_{T}^{b \bar b}$ turns out to be $\sim$ 1 for signal. 
In backgrounds, particularly in top pair production there are 
additional hard jets arising due to the hadronic decay 
of W(since we are giving a veto on the second lepton) 
resulting in $R_{T}^{b \bar b} < 1$. Thus a judicious choice of a upper 
cut on $R_T^{b\bar b}$ suppresses backgrounds enormously without 
affecting the signal much.
\\
$\bullet$ ${\phi^{b\bar{b}}}$:
The azimuthal angle  $\phi^{b\bar{b}}$ is defined as the angle 
between two b-jets in the transverse
plane. In the signal process the angle is 
expected to be small, but in backgrounds, 
for example in top pair production
they are in general widely separated. 
It has to be emphasized here that with the increase in $\N0_{2}$ mass, 
the $h \to b\bar{b}$ system gets more and more boosted and hence the b 
jets become more and more 
collinear which is an ideal situation 
for the jet substructure analysis 
described in the 
following section. We find 
that a reasonable cut on $\phi^{b\bar{b}}$($\phi^{b\bar{b}}\le 2$) 
suppresses the background considerably.
\\
$\bullet$ $ {m_T(\ell,\PMET)}$: The transverse mass is  defined as
$m_T = \sqrt{2 p_T^\ell\PMET(1-\rm cos\phi(\ell,\PMET))}$, 
where $\phi(\ell,\PMET)$
is the azimuthal angle between the lepton and $\PMET$ direction. The value
of $m_T(\ell,\PMET)$ is expected to be restricted by W mass 
if both leptons and $\PMET$ originate
from W decay, which is the case for backgrounds, particularly for
$t \bar t$ and $W b \bar b$ channels. Therefore, a reasonable cut on 
$m_T(\ell,\PMET)$ is found to be extremely effective to reduce the 
background level. 
\\
$\bullet$ $m_{b\bar b}$:
As mentioned above the invariant mass of two b-jets is very useful in
isolating the signal region. In method A, this reconstruction is straight 
forward
and is performed using two b-jets momenta obtained by 
matching b-jets with b-quarks. 
However, in method B, we use jet substructures to find b-jets 
inside a "fat-jet" from the Higgs decay. 
The use of jet substructure for the reconstruction of hadronic decays of 
boosted $W$, $Z$, Higgs boson and top quark has received  considerable 
attention in recent years and the available literature is steadily increasing
\cite{tilman}.
In our present study this method was motivated 
following the work  of Ref.~\cite{butterworth} where the authors 
reconstructed the Higgs mass using jet substructures to  
increase the signal sensitivity.
The efficiency of jet substructure technique depends on the boost factor 
of the decayed object. A highly boosted system
ensures that decay products are well collimated and appear as a "fat-jet". 
However, in the scenario of interest to us Higgs
 is moderately boosted as its $p_T$ depends
on $\Delta m=m_{\N0_2} - m_{\N0_1}$.  
In our analysis we first cluster all the stable final state particles into a 
``fat jets'' using the C/A algorithm~\cite{CA} with R =1.2 as implemented in 
Fastjet~\cite{salam}. We select "fat jets" with $p_T \ge$100~GeV and
$|\eta|<$2.5 and then perform jet substructure analysis. 
There are various methods of finding jet substructures \cite{tilman}.
We use the mass drop(MD) method\cite{butterworth}(coded in the {\tt FastJet} 
package \cite{salam}) in our analysis optimizing the two input parameters, 
$\mu=0.4$ and ${\rm y_{cut}=0.1}$. 
In the simulation we use PYTHIA event generator 
by setting Tune $Z2^*$ 
parameters described in Ref.~\cite{tunez2} for  
underlying event modeling. 
In Figure.\ref{fig:fig1}, we show the reconstructed 
Higgs mass following method A(blue) and B(red) corresponding to parameters P2.
This figure clearly demonstrates the usefulness of the jet substructure 
technique for Higgs mass reconstruction. In case of method A, some of the soft jets 
are incidentally passing the
matching criteria resulting in a spread towards the lower side, whereas in
the jet substructure method this type of contamination 
is avoided by the filtering procedure described in \cite{butterworth}.  

\section{Results}
{{\underline {Method~A:}}}
\begin{table}[t!]
\begin{center}
\begin{tabular}{lllllllll}
\hline
Process & $\sigma$(pb) & $N_{EV}$ & $1\ell $ & $R_T^{b\bar{b}}$ & 
$m_{b\bar{b}}$ 
 & $\PMET$     &$\phi_{bb}$& c.s. \\ 
& & &2b-jets &$\ge$0.7 &110-130 &$\ge$175 &$\le$2 &(fb)  \\
\hline
P1 & 0.175 & 0.1M & 1392 & 1162 & 723  & 92 & 76 & 0.065\\
P2 & 0.065 & 0.1M & 1767 & 1478 & 933  & 217 & 178 & 0.06 \\ 
P3 & 0.03 & 0.1M  & 2142 & 1774 & 1122 & 424 & 391 & 0.055 \\
\hline  
Wh & 0.58 & 50K & 702 & 594 & 394 & 8 & 2 & 0.01        \\
Zh & 0.3 & 50K & 210 & 162 & 51 & 1 & 1 & 0.003         \\
$Wb\bar{b}$ & 3 & 619685 & 26841 & 24513 & 2269 & 8 & 3 & 0.014 \\
$Zb\bar{b}$ & 5.1 & 378098 & 3863 & 2937 & 269 & $\rm <1 $ &$\rm <1 $ & $\rm <1 $    \\
$t\bar t$ & & & & & & & &  \\  
5-100 & 48.2 & 4M & 207335 & 94337 & 10145 & $\rm <1 $ & $\rm <1 $  & $\rm <1 $ \\
100-200 & 36.3 & 2M & 158450 & 50967 & 1205 & 8 & 1 & 0.01 \\
200-500 & 9.5 & 1M & 134238 & 22473 &116 & 42 & 4 & 0.02 \\
\hline
\end{tabular} 
\caption{\label{table:table2} Event summary for signal and backgrounds(method A)
for 8~TeV after each set
of cuts described in the text. The $t \bar t$ 
events are simulated for different $\hat p_T$ bins as shown. Efficiency 
for tagging two b jets is multiplied in the last column.
Note that for entries with $\rm < 1$  the event yield
in our case is  0;
 however since we have simulated finite number 
of events we denote them as
 $\rm < 1$.
  The energy units are
in GeV.}
\end{center}
\end{table}

\begin{table}[t!]
\begin{center}
\begin{tabular}{lllllllll}
\hline
Process & $\sigma$(pb) & $N_{EV}$ & $1\ell $ & $R_T^{b\bar{b}}$ & 
$m_{b\bar{b}}$ 
 & $\PMET$     &$\phi_{bb}$& c.s. \\ 
& & &2b-jets &$\ge$0.7 &110-130 &$\ge$175 &$\le$2 &(fb)  \\
\hline
P1 & 0.502 & 0.1M  & 3867 & 2500 & 1213 & 89 & 73 & 0.18  \\
P2 & 0.202 & 0.1M & 4391  & 2756  & 1381  & 273 & 229  & 0.23 \\ 
P3 & 0.104 & 0.1M  & 4517 & 2824 & 1431 & 373 & 323  & 0.17 \\
\hline  
Wh & 1.26 & 0.1M & 3002 & 1639 & 750  & 21 & 15  &0.09 \\
Zh & 0.69 & 0.1 M & 799 & 280 & 85  & 1 & 1  &0.004 \\
$Wb\bar{b}$ & 4.5 & 362018  & 57764 & 47883 & 44160 & 3948 & 9 & 0.055 \\
$Zb\bar{b}$ & 7.2 & 406110 & 442 & 380 & 322 & $\rm <1 $ & $\rm <1 $ & $\rm < 1$    \\
$t\bar t$ & & & & & & & &  \\  
5-100 & 188  & 10M & 1163903  & 188856   & 16910 & 7  &  4 & 0.04  \\
100-200 &  156 & 10M & 1202319 & 82970 & 4367 & 70 & 32 & 0.25 \\
200-500 & 48.5 & 1M& 133840   & 2020 &252 & 61 & 32 & 0.8  \\
\hline
\end{tabular} 
\caption{\label{table:table3}
Same as Table 2 but for 14 TeV.
The same conventions as in 
Table \ref{table:table2} are used.
 }
\end{center}
\end{table}

In this section we discuss the simulation strategy of  
signal and backgrounds by reconstructing the Higgs mass 
out of two identified b-jets obtained by matching 
techniques as discussed above.
In order to eliminate SM backgrounds additional cuts  
are applied with the following requirements:
\\
$\bul ~{\rm R_T^{b\bar b}}\ge$0.7,\\
$\bul ~{\rm m_{b\bar b}}$=110-130 GeV,\\
$\bul ~ \PMET\ge$175~GeV,\\
$\bul ~ {\rm \phi_{b\bar b}}\le$2. \\
In Table \ref{table:table2} we present event summaries of signal
for three benchmark points  shown in Table  \ref{table:table1}, 
along with backgrounds  
after applying these set of cuts. The second and 
third column present the raw leading order(LO) 
cross section and number of events simulated respectively. 
In the
fourth column, we present the number of events requiring one single hard lepton
along with two identified b-jets and veto the second lepton as well. Although
we simulated all possible SM backgrounds including QCD, $tb$, $tbW$, 
but we present results  
only for non-negligible contributing channels. 
It clearly demonstrates that the $R_T^{b\bar b}$ cut is very effective in
reducing backgrounds by an enormous amount, but 
except for channels, like $Wb\bar{b}$ and $Wh$. Selection of events 
in the Higgs mass window between 110-130 GeV is also useful to remove 
backgrounds keeping 
almost more than 50\% of signal events.  
Finally, a very strong $\PMET$ cut is used to eliminate remaining backgrounds,
but at the cost of a sizable signal cross section; 
nevertheless we retain a good number of signal events. 
After all cuts, we find the 
total background cross section is
about 0.057 fb with dominant contribution from $t \bar t$,
whereas signal cross sections are in the range 0.065-0.055 fb. In both 
cases we use LO production cross sections.
However, if we use NLO cross sections by multiplying K-factors which is 
$\sim$ 1.5 for signal~\cite{prospino} and about 1.6 for $t \bar t$
\cite{kidonakis}, then assuming a luminosity 100~ $\lumi$, one can expect 
S/$\sqrt{B}$ about 3.5 for these mass ranges of $\CH_1$ and $\N0_2$.

In Table 3 , we present results for 14 TeV energy
corresponding to the same set of benchmark points along with the SM
background. We observe that 
the signal efficiency remains fairly the same as 8~TeV
with the enhancement occurring only due to the increase in cross section.
  The top background however increases significantly 
due to a presence of a stronger
 missing momentum and more reconstruction of 
Higgs mass from the $b\bar{b}$ system.
 The total background cross section 
at 14~TeV turns out to be 1.23 fb as compared to the signal cross 
sections
which are between 0.18 fb and 0.25 fb. As a consequence
 it becomes
difficult to observe a signal with low luminosity options
in this approach. 
However for an integrated luminosity of $\rm 1000 fb^{-1}$  it
may be possible to 
observe a signal in this method at a 5$\rm \sigma$ level.

\begin{table}[H]
\begin{center}
\begin{tabular}{lllllllll}
\hline
Process & C.S. & $N_{EV}$ & $m_{b\bar{b}}$ & Lepton 
& $m_T(\ell,\PMET)$ & 
$R_T^{b\bar{b}}$ & $\PMET$ & C.S. \\ 
 &(pb) & & &$\ge$20 &$\ge$ 90 &$\ge$0.9 &$\ge$ 125  & (fb) \\
\hline
P1 & 0.175 & 25K & 179 & 55 & 24 & 19 & 12 & 0.042\\
P2 & 0.065  & 10K & 168 & 42 & 23 & 18 & 12 & 0.04 \\ 
P3 & 0.03  & 10K & 273 & 75 & 44 & 36 & 31 & 0.045 \\
\hline  
Wh & 0.58 & 0.1M & 871 & 239 & 12 & 11 &$\rm < 1$ & $\rm < 1 $ \\
Zh & 0.3 &  0.2M & 1698 & 37 & $\rm <1$ & $\rm <1$ & $\rm <1$ & $\rm < 1$ \\
$Wb\bar{b}$ & 3 & 619671 & 191 & 111 & 10 & 8 & $\rm <1$ & $\rm < 1$ \\
$Zb\bar{b}$ & 5.1 & 378086 & 81 & 13 & $\rm <1$ & $\rm <1$ & $\rm <1$ & $\rm < 1$ \\
$t\bar{t}$ & & & & & & & \\
5-100 & 48.2 & 5M  & 1669 & 454 & 38 & 1 & $\rm < 1$ & $\rm < 1$ \\
100-200 & 36.3 & 4M & 1583 & 440 & 42 & 3 & 1 & 0.005 \\
200-500 & 9.5 & 1M & 315 & 98 & 9 & 2 & $\rm < 1$ & $\rm < 1$\\  
\hline
\end{tabular} 
\caption{\label{table:table4} Event summary for signal and
 backgrounds(method B) for 8~TeV after each set
of cuts described in the text. The same conventions as in 
Table \ref{table:table2} are used.}
\end{center}
\end{table}

{{\underline {Method~B}}:
In this method we apply jet substructure technique in reconstructing
mass of Higgs within the mass window between 117 - 128 GeV and  
with additional cuts 
as before to control background events,
\\
$\bul~ {\rm m_T(\ell,\PMET)} \ge$ 90~GeV
\footnote{Note  that we have taken 
the finite width effects of W boson into account
in our simulation. 
This results in a  tail in the ${\rm m_T(\ell,\PMET)}$ 
distribution in processes like Wh and t$\rm \bar{t}$.
Thus forced us to opt for a higher 
value for ${\rm m_T(\ell,\PMET)}$  selection cut.},\\
$\bul~ {\rm R_T^{b\bar{b}}}\ge$0.9,\\
$\bul~ \PMET\ge$ 125 GeV(150~ GeV for 14 TeV).
\\ 
After the Higgs mass reconstruction the remaining
stable particles are used to find jets with C/A algorithm 
with $\Delta R=0.5$, $p_{T}\ge50$ GeV, $|\eta|\le 2.5$.
The Table \ref{table:table4} displays the robustness of 
$R_T^{b\bar{b}}$ cut along with $m_T(\ell,\PMET)$ leading to a suppression of
backgrounds to a negligible level without affecting signal 
significantly. 

\begin{table}[H]
\begin{center}
\begin{tabular}{lllllllll}
\hline
Process & C.S. & $N_{EV}$ & $m_{b\bar{b}}$ & Lepton 
& $m_T(\ell,\PMET)$ & 
$R_T^{b\bar{b}}$ & $\PMET$ & C.S. \\ 
 &(pb) & & &$\ge$20 &$\ge$ 90 &$\ge$0.9 &$\ge$ 150  & (fb) \\
\hline
P1 & 504 & 25K & 242 & 55 & 23 & 16 & 5 & 0.05 \\
P2 & 204  & 25K & 461 & 113 & 55 & 43 & 26 & 0.1 \\ 
P3 & 104  & 25K & 713 & 197 & 116 & 67 & 46 & 0.095  \\
\hline  
Wh & 1.3 & 0.1M & 946 & 289 & 17 & 11 & 4 & 0.026 \\
Zh & 704 & 0.1M  & 866 & 13 & 1 & $\rm < 1$ & $\rm < 1$ & $\rm < 1$  \\
$Wb\bar{b}$ & 5.5 & 431062 & 159 & 92 & 8 & 7 & $\rm < 1$ & $\rm < 1$\\
$Zb\bar{b}$ & 7.2 & 571166 & 150 &$\rm < 1$ & $\rm < 1$ & $\rm < 1$ & $\rm < 1$ & $\rm < 1$ \\
$t\bar{t}$ & & & & & & & \\
5-100 & 190 & 10M  & 4178 & 1016 & 121 & 12 & $\rm < 1$ & $\rm < 1$ \\
100-200 & 158 & 1M & 4463 & 1181 & 137 & 11 & 2 & 0.01  \\
200-500 & 49 & 0.25M & 867 & 296 & 25 & 7 & 1 & 0.02 \\  
\hline
\end{tabular} 
\caption{\label{table:table5} Same as Table 4 
 but for for 14~TeV.
 The same conventions as in 
Table \ref{table:table2} are used.}
\end{center}
\end{table}
Notice that after cuts signal cross sections 
remain the same for all cases although production cross sections 
decrease with the increase of gaugino masses, which is 
compensated by the increase of acceptance efficiencies.
The total background cross section turn out to be 0.007~fb, an order of 
magnitude less than the method A whereas signal cross sections are of
the same level.
Assuming 100~$\lumi$ luminosity, one can expect signal to
background ratio S/$\sqrt{B}$ $\sim$ 7 using NLO cross sections as before.
It implies that probing the Higgs signal in this 
channel is promising with 8 TeV  LHC energy and high luminosity options.
In both cases signal sensitivity is low because of the tiny production
cross section in comparison with the backgrounds.

For 14~TeV energy, as presented in  
 Table \ref{table:table5}, we find that the results are not significantly 
different for method B. Comparing 
Table \ref{table:table4} and Table \ref{table:table5} we
observe a better reconstruction of the Higgs mass because of 
the enhanced boost of the  $b\bar{b}$  system at 14 TeV
energy. However this gain is diluted  due to an 
increase in $\PMET$  cut compared to 8 TeV to suppress the  
backgrounds. It has to be noted that at 14 TeV 
we receive a finite background contribution from Wh process
due to an increase in $\PMET$.   
We find that after all cuts the total background cross section
 is 0.05 fb while the signal cross sections vary between
0.05 fb  to 0.1 fb . It is therefore possible to discover
a signal for this type of parameter space
 at the 5 $\rm \sigma$ level
 at $\sim $ 100 $fb^{-1}$ luminosity.

\section{Summary}
We investigate the discovery potential of a Higgs signal  
in $\CH_1\N0_2$ production
and its subsequent decay channels at 8 TeV and 14 TeV LHC energy.
This study is performed in the context of
mSUGRA model taking into account the current Higgs mass constraints  
predicted by recent measurements by CMS and ATLAS 
experiments. We simulate signal events in the final state with  
a single hard lepton and $\PMET$ along with 
a reconstructed Higgs mass. The Higgs mass reconstruction is performed
following two ways, first by identifying b-jets using matching technique
(method A) and 
 secondly by using the method of jet substructures (method B).
We present results for both cases and find that the 
latter method is more promising than the former one.
Incidentally, for low luminosity (${\mathcal L} \sim 20fb^{-1}$),
which is the projected luminosity for 8 TeV LHC run, the signal 
cross section is 
too low to be observed and hence we require high luminosity.   
For instance, we expect a significance of about $\sim$ 7 for  
250-400 GeV masses of $\CH_1$ and $\N0_2$ and $m_h \sim$125 GeV with 
100~$\lumi$ luminosity
by using jet substructure method for 8~TeV. We also
performed the analysis for 14 TeV energy and found identical results.
The observations made in this paper therefore suggest
that the jet substructure method works better for both 8 TeV and 14 TeV
LHC energy.  
It is observed that results do not change significantly for other values
of $\tan\beta$.
In order to increase the sensitivity of Higgs signal one needs to devise more
effective selection cuts to isolate tiny signal events out of the huge 
backgrounds. 
If Higgs is discovered it is worthwhile to study this
channel to identify the model  framework.
The signal acceptance efficiency is  dependent on $\Delta m$, 
which is sensitive to different models. Therefore,  
our conclusions are 
model specific and expected to be different in the case of other SUSY 
models, particularly in models where mass relations among gauginos follow a 
different pattern. It is also worth investigating the feasibility of 
detecting supersymmetric Higgs in the ${\rm h \to \tau^+ \tau^-}$ channel which 
will be presented in a future work \cite{diptimoy}.

\end{document}